\documentclass[journal=jpccck,manuscript=article,hyperref=true,doi=true,keywords=true,abbreviations=true]{achemso}

\usepackage{amsmath,amssymb}
\usepackage{subcaption}   
\usepackage{placeins}     
\usepackage{textcomp}
\usepackage{xcolor}       
\usepackage{ulem}         

\hypersetup{
  citecolor=blue,
  linkcolor=blue,
  urlcolor=blue
}

\title{ Interplay between Isomerization and Spin Crossover in 1D Fe-Indigo Coordination Polymers on Ag substrates}

\author{Ritam Chakraborty}
\affiliation{Theoretical Sciences Unit, Jawaharlal Nehru center for Advanced Scientific Research Jakkur, Bangalore, 560054, India}
\author{Hongxiang Xu}
\affiliation{Technical University of Munich, TUM School of Natural Sciences, Physics Department E20, 85748 Garching, Germany}
\affiliation{School of Electronics, Peking University, Beijing, 100871, China}
\author{Biao Yang}
\affiliation{Technical University of Munich, TUM School of Natural Sciences, Physics Department E20, 85748 Garching, Germany}
\affiliation{State Key Laboratory of Bioinspired Interfacial Materials Science, Institute of Functional Nano and Soft Materials (FUNSOM), Soochow University, Suzhou, 215123, P.~R.~China }
\author{Harshdeep Singh Chhabra}
\affiliation{School of Physics, Indian Institute of Science Education and Research Thiruvananthapuram,
Thiruvananthapuram, 695551, Kerala, India}
\author{Joachim Reichert}
\affiliation{Technical University of Munich, TUM School of Natural Sciences, Physics Department E20, 85748 Garching, Germany}
\author{Johannes V. Barth}
\affiliation{Technical University of Munich, TUM School of Natural Sciences, Physics Department E20, 85748 Garching, Germany}
\author{Anthoula C. Papageorgiou}
\affiliation{Technical University of Munich, TUM School of Natural Sciences, Physics Department E20, 85748 Garching, Germany}
\affiliation{Laboratory of Physical Chemistry, Department of Chemistry, National and Kapodistrian University of Athens, 157~71 Athens, Greece}
\author{Shobhana Narasimhan}
\affiliation{Theoretical Sciences Unit, Jawaharlal Nehru center for Advanced Scientific Research Jakkur, Bangalore, 560054, India}
\affiliation{School of Advanced Materials, Jawaharlal Nehru center for Advanced Scientific Research Jakkur, Bangalore, 560054, India; International center for Theoretical Sciences, Shivakote, Hesaraghatta Hobli, Bengaluru, 560080, India}
\email{shobhana@jncasr.ac.in}

\abbreviations{SCO,DFT,CP,STM,ncAFM,XPS,PDOS,LS,HS,FM,PAW,BFGS}
\keywords{spin crossover, coordination polymers, density functional theory, on-surface synthesis, indigo, iron}

\begin{document}

\begin{abstract}
Spin-crossover (SCO) compounds offer a route to switchable molecular
functionality in reduced dimensions. However,
one-dimensional (1D) SCO chains, which offer the possibility to study ligand fields other than the paradigmatic octahedral field, remain comparatively little studied. 
Here, we use first-principles density functional theory (DFT+$U$) to
investigate Fe-indigo coordination-polymer chains synthesized experimentally on
Ag(111) and Ag(100) substrates. These display a rich interplay between changes in ligand field (isomerization) and spin crossover. On-surface isomerization on Ag(111)
interconverts (N,O)-chelated \textit{trans} configuration and 
(N,N)-/(O,O)-chelated \textit{cis} configurations at the Fe centers.
 The lowest-energy
\textit{trans} and \textit{cis} solutions on Ag(111) have different
spin configurations over the  interval
$0.66<U<3.00$~eV. At the reference value $U=1$~eV, the preferred
\textit{trans} solution is the mixed LS--LS--HS configuration, whereas
the preferred \textit{cis} solution is LS--LS--LS. The
experimentally observed preference for \textit{cis} chains on Ag(111)
and \textit{trans} chains on Ag(100) is reproduced for the 
range $0.88<U<3.75$~eV. 
  To interpret these results, toy models and spin-resolved Fe
$3d$ projected densities of states are used, while freestanding-chain calculations reveal a
strain-sensitive LS--HS competition. These results provide a
microscopic explanation for isomerization-controlled spin-state switching in
a 1D coordination polymer.
\end{abstract}

\maketitle

\section{Introduction}

Spin-crossover (SCO) systems can switch between low-spin (LS) and
high-spin (HS) spin states through the competition
between ligand-field splitting and intra-atomic spin-pairing
energy.\cite{gutlich2004,halcrow2011} The transition is accompanied
by changes in magnetic moment, electronic structure, and molecular
geometry, making SCO materials attractive for a variety of applications including molecular switching,
sensing, and spintronics.\cite{halcrow2013,kipgen2021,li2018}
Because this energetic balance is sensitive to structural confinement
and molecule-substrate interactions, reduced-dimensional SCO systems
provide an opportunity to control spin states through their local
coordination environment.\cite{halcrow2008,sanchezdearmas2023,rohlf2019,thakur2021}

For applications, SCO molecules are usually deposited on a substrate. Surface-supported SCO molecules \cite{fourmental2019importance, bellec2026interplay} and one-dimensional coordination
polymer (CP) chains are also especially appealing because their structure
and spin state can, in principle, be addressed at the molecular scale.
On-surface coordination has enabled studies of magnetic anisotropy,
lattice-directed spin-state phase selection, collective spin switching,
and cooperative SCO dynamics in individual chains.\cite{umbach2012,liu2021,liu2020,denawi2018}
Silver surfaces in particular have proven a versatile platform for such
studies, hosting spin-crossover behaviour in both Fe(II) and Co(II)
complexes.\cite{zhang2019,johannsen2022} First-principles calculations
provide a complementary microscopic view of the spin-state energetics
and metal--ligand orbital interactions, particularly when hybridization
with a metallic substrate complicates direct experimental
interpretation.

{In most on-surface spin-crossover studies, the switching functionality is intrinsic to a pre-formed molecular complex: the metal center is already fully coordinated before deposition, and interaction with the substrate modifies a spin transition that exists in the isolated complex.\cite{sanchezdearmas2023,rohlf2019,thakur2021} A different situation arises in surface-supported coordination polymers, where competing spin states emerge only after metal--ligand assembly. Examples include Ni centers coordinated by deprotonated tetrahydroxybenzene on Au surfaces,\cite{liu2020,liu2021} and Fe--quinone chains on Au(110), for which spin-crossover behaviour has been predicted by DFT+$U$ calculations.\cite{denawi2018} In these systems, however, the local coordination environment of the metal remains essentially fixed, while the spin state is controlled by external parameters such as substrate-imposed metal--metal spacing or tunnelling electrons.} 

Here, we study a related mechanism in the Fe--indigo coordination polymer synthesized on Ag surfaces by Xu \textit{et al.}\cite{xu2024} Neither constituent is independently spin-crossover-active: isolated Fe adatoms do not possess a molecular ligand field, while indigo contains no metal center. The relevant Fe spin states therefore arise only after formation of the coordination polymer. A further distinction is that the polymer can undergo a structural transformation that changes the local symmetry of the ligand field at the Fe centers. Indigo is well suited to such coordination rearrangements because its deprotonated form supports several metal-binding motifs, including (N,O)-chelation.\cite{kaim2019,beck2020} Coordination-induced dehydrogenation makes the central C--C bond rotatable, enabling \textit{trans}-to-\textit{cis} isomerization (see Fig.~1). This transformation changes the Fe coordination from an (N,O)-chelated environment in the \textit{trans} chain to distinct (N,N)- and (O,O)-chelated environments in the \textit{cis} chain. The two structures therefore generate different ligand fields and different Fe $3d$-orbital level splittings. The key distinction of the present system is therefore that the spin-state change is coupled to an intrinsic reorganization of the Fe coordination environment through on-surface \textit{trans}--\textit{cis} isomerization, rather than to an externally imposed structural or electronic perturbation.

In our previous work \cite{xu2024},  Fe atoms and indigo molecules were co-deposited on Ag(111) and Ag(100), and the resulting one-dimensional Fe--indigo coordination polymers were characterized using scanning tunnelling microscopy, bond-resolved non-contact atomic force microscopy, X-ray photoelectron spectroscopy, and density functional theory (DFT). Thermal annealing induced \textit{trans}-to-\textit{cis} isomerization on Ag(111), accompanied by a change in Fe--ligand coordination, whereas the corresponding transformation was not observed on Ag(100). Here, we use DFT+$U$ calculations to determine whether this structural change also modifies the preferred Fe spin state. Two self-consistent spin states are obtained, denoted LS and HS and corresponding nominally to $S=1$ and $S=2$, respectively. Calculations for the Ag(111)- and Ag(100)-supported chains, together with simplified coordination models and freestanding-chain calculations, are used to separate the effects of coordination geometry, substrate hybridization, and strain. We show that this isomerization-induced change in ligand field symmetry is sufficient to alter the relative stability of the Fe spin states, thereby establishing a direct link between structural isomerization, ligand-field symmetry, and spin-state selection in the coordination polymer.

\section{Computational Details}

All calculations were performed using spin-polarized density functional
theory as implemented in the Quantum ESPRESSO suite of
codes.\cite{giannozzi2009quantum} { Calculations were restricted to collinear spin configurations only.} Exchange and correlation were treated
with the nonlocal vdW-DF2-B86R functional.\cite{hamada2014van,lee2010higher}

The Fe $3d$ states were treated within DFT+$U$, following Dudarev's
formulation\cite{dudarev1998electron} and its implementation by
Cococcioni and de Gironcoli.\cite{cococcioni2005linear} It is well-known that while results can be quite sensitive to the choice of the Hubbard parameter $U$, the determination of an appropriate value of $U$ remains a tricky question. Even computing the value of $U$ from first principles does not necessarily lead to physically correct results.\cite{Lu2014, Zhang2023} We have therefore varied $U$ from 0 to 4~eV (in steps of 1 eV); we will later compare calculated results with experimental data, so as to finally settle on a value for $U$.

Projector augmented-wave (PAW)
pseudopotentials were used,\cite{blochl1994projector} with plane-wave
cutoffs of 60~Ry and 300~Ry for the wavefunctions and charge density,
respectively. Marzari--Vanderbilt cold smearing\cite{marzari1999thermal}
with a width of 0.005~Ry was used to aid convergence.

The Ag(111) and Ag(100) surfaces were represented by three-layer slabs
separated from their periodic images by more than 18~\AA{} of vacuum
along the surface normal. For Ag(111), the coordination polymer was
modelled using the surface supercell
$\begin{pmatrix}7&1\\4&8\end{pmatrix}$, containing three
dehydrogenated indigo molecules and three Fe atoms. The Ag(100) system
used the surface supercell $\begin{pmatrix}5&-4\\1&2\end{pmatrix}$, { containing one dehydrogenated indigo molecule and one Fe atom}. {These choices of supercell were guided by STM data, as has been previously discussed.\cite{xu2024}

The Fe-indigo overlayer and the upper two Ag layers were relaxed using
the Broyden-Fletcher-Goldfarb-Shanno (BFGS) algorithm\cite{broyden1973local,Fletcher_1970,goldfarb1970family,shanno1970conditioning}
until every force component was below $10^{-3}$~Ry/Bohr; the coordinates of atoms in the bottommost Ag
layer was kept fixed. Geometry optimizations on Ag(111) used  $\Gamma$
point sampling of the supercell Brillouin zone, whereas Ag(100) calculations used a $1\times10\times1$
Monkhorst--Pack grid.\cite{monkhorst1976special} The optimized bulk Ag
lattice constant was 4.11~\AA, close to the experimental value of
4.09~\AA; the corresponding nearest-neighbour spacing in the Ag(111)
plane is $4.11/\sqrt{2}=2.91$~\AA. 

Throughout the manuscript, LS and HS denote the two converged local Fe
spin manifolds, nominally $S=1$ and $S=2$, respectively. 

{The three-unit supercell used for Ag(111) allows us to sample both uniform-spin (e.g., LS-LS-LS) and mixed-spin (e.g., LS-LS-HS) configurations. For Ag(100), we restrict ourselves to the one-unit supercell, and thus consider only uniform-spin configurations.  To summarize the calculations performed on the on-surface CPs,
for Ag(111), we have considered LS-LS-LS, LS-LS-HS, LS-HS-HS and HS-HS-HS spin configurations in both {\it{trans}} and {\it{cis}} conformations, using five different values of the Hubbard parameter $U$. Thus, we have performed a total of $ 4 \times 2 \times 5 = 40$ calculations on Ag(111). Similarly, on Ag(100) we have performed $2 \times 2 \times 5 = 20$ calculations. We note that for each system, only one spin configuration will correspond to the ground state; converging the system instead into another metastable spin state can frequently be a challenging task. Despite taking the utmost care, we were unable to converge the system to mixed-spin cis CPs on Ag(111), suggesting that they are not even local minima in spin space.}

\section{Results and Discussion}

Before discussing the Fe spin-state energetics, it is useful to recall
why the \textit{trans}-to-\textit{cis} transformation becomes possible
in the coordination polymer. Free indigo strongly favors the
\textit{trans} form, in part because an intramolecular N--H$\cdots$O
hydrogen bond that is absent in the \textit{cis} isomer stabilizes it by
about 0.79~eV.\cite{xu2024}  Fe-coordination-induced dehydrogenation removes this
hydrogen bond and converts the central C$=$C bond into a rotatable
single bond. In the stepwise reaction pathway analyzed previously,
subsequent adsorption and Fe coordination progressively reduce and
finally reverse the \textit{trans} vs.~\textit{cis} energy difference; in
the fully coordinated chain on Ag(111), we have found that the \textit{cis} conformation is lower in energy than the {\it{trans}} conformation
 by about 91~meV per unit.\cite{xu2024} 

\subsection{Fe atoms in $D_{2h}$ and $C_{2v}$ symmetry ligand fields}
As already mentioned, the majority of SCO systems that have been studied till date contain a central metal atom (usually Fe) in an octahedral ligand field. The five $d$ orbitals then split into $e_g$ and $t_{2g}$ manifolds, as is well known. A much less studied system is a square planar ligand field with $D_{4h}$ symmetry.\cite{Kawakami2009} In this case, the five $d$ orbitals split into four levels, with only $d_{xz}$ and $d_{yz}$ remaining degenerate. The LS and HS states correspond to $S =1$ and $S=2$, respectively.

Here, we wish to lower the symmetry  further still, to either $D_{2h}$ or $C_{2v}$. Since such systems have been little studied, it is worth recalling what happens to the $d$ orbitals in such cases. Figure S2  in the SI shows the character tables and irreducible representations for both $D_{2h}$ and $C_{2v}$. In both cases, all five $d$ orbitals become non-degenerate. The exact ordering of the energies of these $d$ orbitals will be highly system-dependent, and different for the two cases.

As an illustrative example, we first consider two hypothetical toy model systems. They are shown in Fig.~S1 in the SI. The systems consist of an FeN$_2$O$_{2}$H$_6$ moiety, in either {\it{trans}} ($D_{2h}$) or {\it{cis}} ($C_{2v}$) conformation. We have performed non-spin-polarized DFT calculations on these conformations; this was done so as to examine the effects of different symmetries alone, not yet considering the effects of different possible spin-state occupation schemes. The calculations confirm the lifting of $d$-orbital degeneracy and the changes in ordering of the different orbitals (see Fig.~S3 in SI). For the {\it{trans}} configuration, the
ordering of energies (in increasing order of energy) is found to be $d_{xz}$, $d_{z^2}$
$d_{xy}$, $d_{yz}$, and $d_{x^2-y^2}$,  whereas for the {\it{cis}} conformation it is found to be $d_{z^2}$, $d_{xy}$
$d_{yz}$, $d_{xz}$, and $d_{x^2-y^2}$. When spin polarization is introduced, the LS and HS states correspond again to $S=1$ and $S=2$, as was the case for square planar symmetry.

\subsection{Spin-state and conformer energetics on Ag(111) and Ag(100): DFT Results as a function of $U$.}

We first perform DFT calculations to find the total energy and the geometric structure as the Hubbard parameter $U$ is varied. On the Ag(111) or Ag(100) substrates, the Fe atoms are in ligand fields that are close to ${D_{2h}}$ or $C_{4v}$, though slight deviations in symmetry occur because of substrate effects.

As examples, the lowest energy relaxed structures, obtained using $U =1$ eV, of the {\it{trans}} and {\it{cis}} CPs formed on Ag(111) are shown in Fig.~\ref{fig:CP-on-Ag111}.  These geometries correspond to those in the lowest-energy spin state for each of these two conformations, viz., LS-LS-HS for {\it{trans}} and LS-LS-LS for {\it{cis}}. Changing the spin state results in slight but important changes in geometry, especially in the lengths of Fe-N and Fe-O bonds.  

\begin{figure}[htbp]
    \centering
    \begin{subfigure}[b]{0.475\linewidth}
        \centering
        \includegraphics[width=\linewidth]{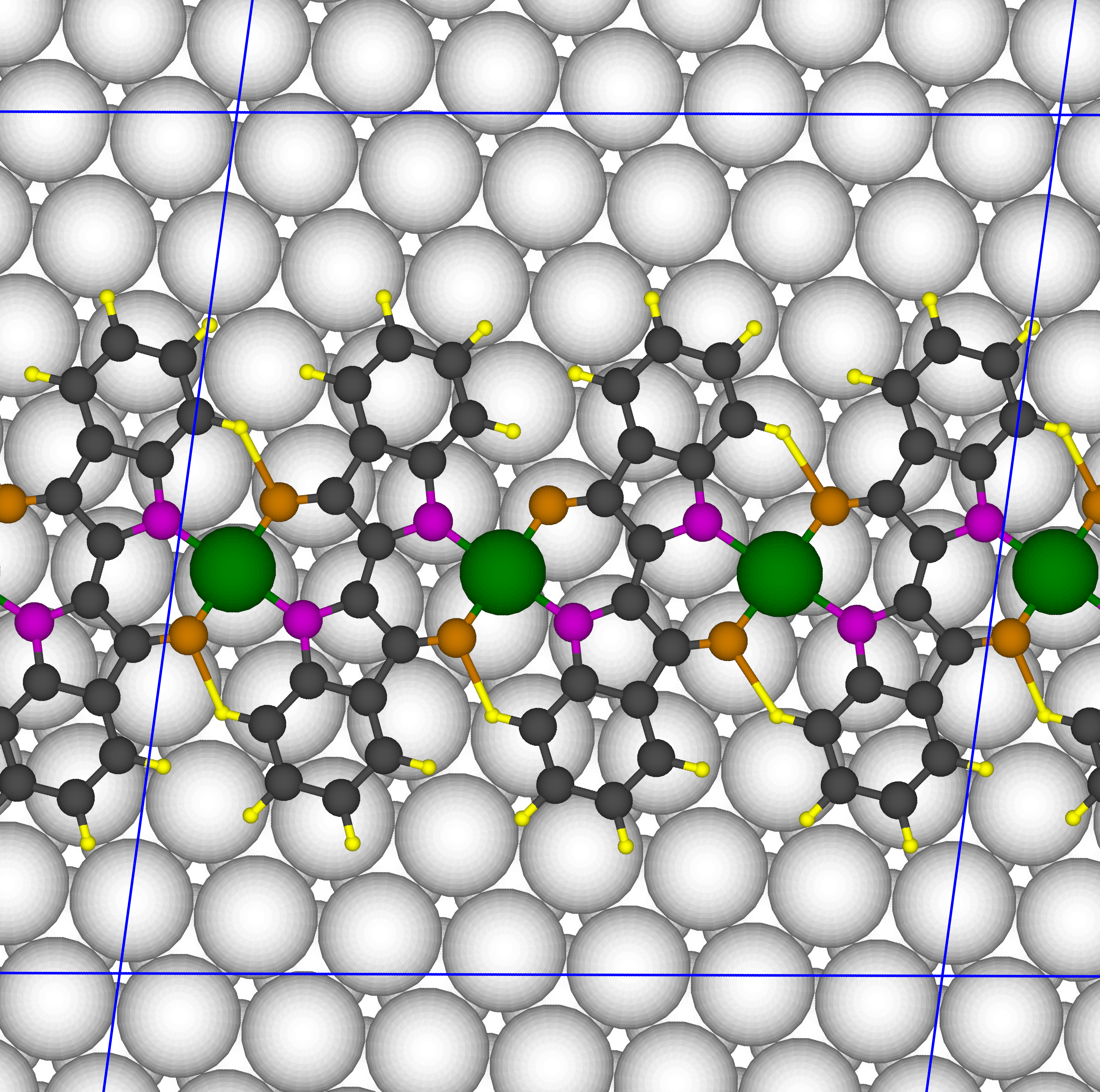}
        \caption{\textit{trans} CP}
    \end{subfigure}
    \hfill
    \begin{subfigure}[b]{0.48\linewidth}
        \centering
        \includegraphics[width=\linewidth]{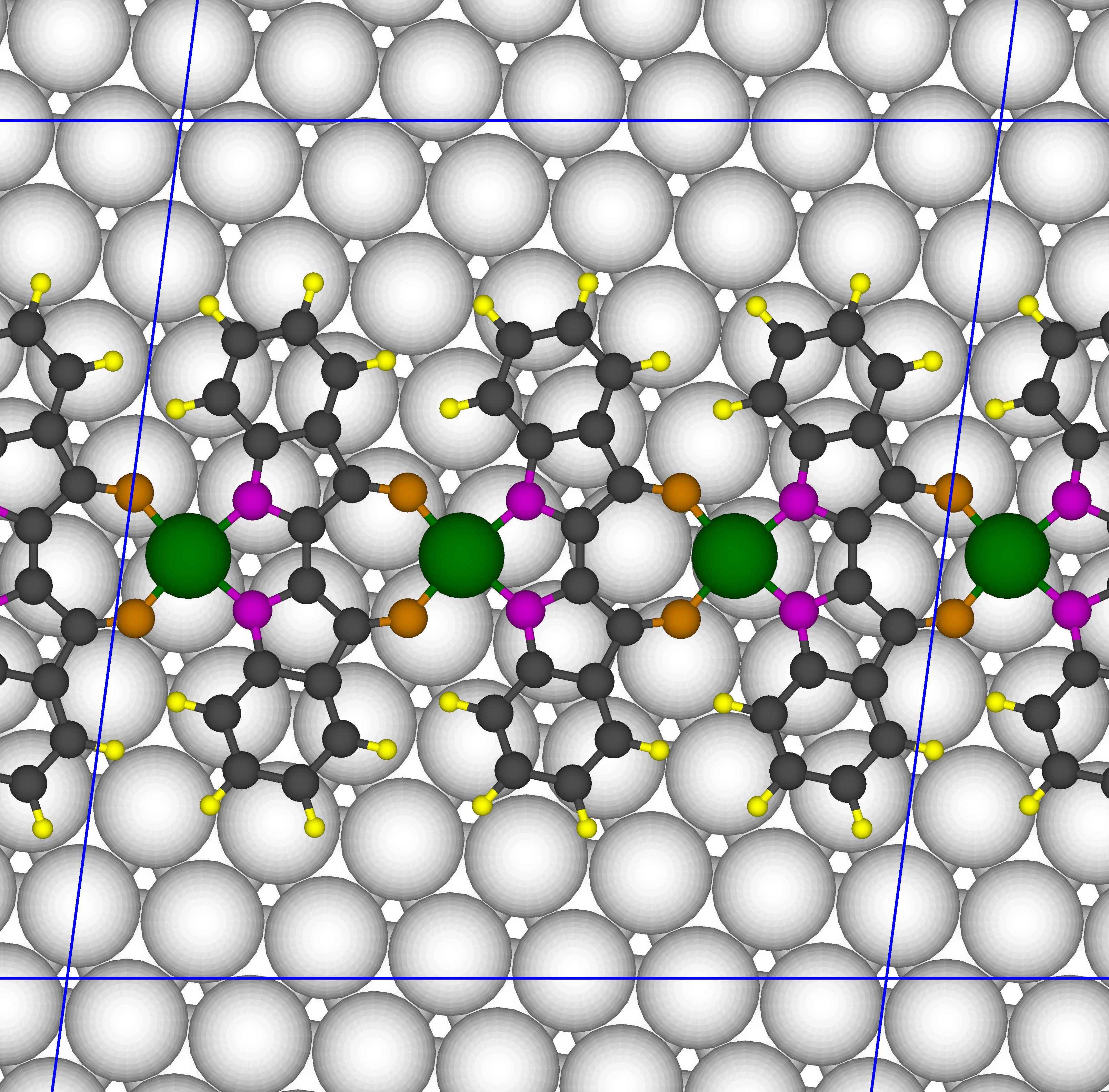}
        \caption{\textit{cis} CP}
    \end{subfigure}
    \caption{DFT optimized structures of the lowest energy spin state of \textit{trans} and \textit{cis} coordination polymers on Ag(111) substrate calculated with $U=1$ eV. Each unit cell
    contains three dehydrogenated indigo molecules and three Fe atoms.
    The structural change modifies the first coordination shell of Fe:
    the \textit{trans} chain is N,O-chelated, whereas the \textit{cis}
    chain contains (N,N)- and (O,O)-chelated Fe environments. Black, yellow,
    pink, orange, green, and white spheres denote C, H, N, O, Fe, and Ag,
    respectively; blue lines mark the surface unit-cell boundary.}
    \label{fig:CP-on-Ag111}
\end{figure}

 As mentioned above, we first computed results over $0 \leq U \leq 4$ eV; we will later use comparison with experiment\cite{xu2024} to identify a physically appropriate value within this range. Figure~\ref{fig:Etot-vs-U-combined} shows
the resulting conformer and spin-configuration energetics. For notational compactness, we will refer to a given combination of spin state + structural conformation as a `phase'. Increasing
$U$ progressively favors HS phases, as is to be expected..

\begin{figure}[htbp]
    \centering
    \begin{subfigure}[b]{0.525\linewidth}
        \centering
        \includegraphics[width=\linewidth]{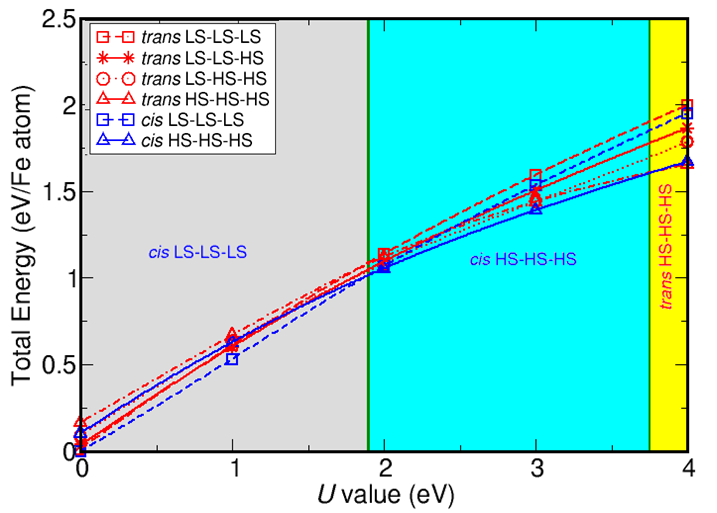}
        \caption{Ag(111)}
        \label{fig:Etot-vs-U-ag111}
    \end{subfigure}
    \hfill
    \begin{subfigure}[b]{0.45\linewidth}
        \centering
        \includegraphics[width=\linewidth]{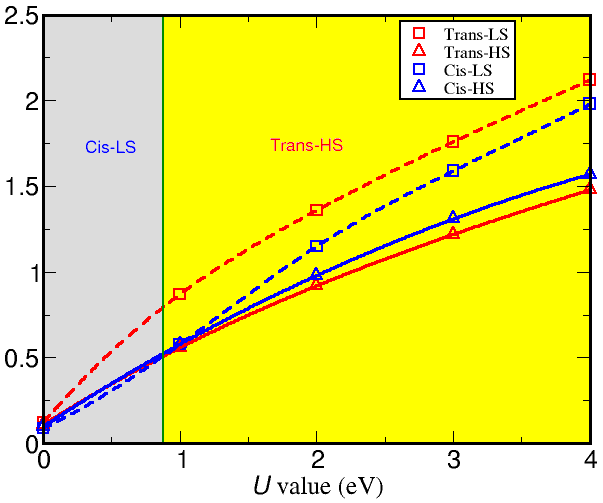}
        \caption{Ag(100)}
        \label{fig:Etot-vs-U-ag100}
    \end{subfigure}
    \caption{Total energies of the \textit{trans} and \textit{cis}
    coordination polymers as a function of $U$ within the
    ferromagnetically aligned spin sector. For readability, the raw
    total-energy scale has been shifted by subtracting the integer part
    of the total energy; this constant vertical offset does not affect
    energy differences or crossover points. Background shading identifies
    the lowest-energy configuration among the plotted self-consistent
    solutions in each $U$ interval. On Ag(111), mixed-spin solutions are
    obtained for the \textit{trans} chain, whereas the converged
    \textit{cis} solutions considered here are all-LS or all-HS. The
    vertical boundaries are obtained by interpolation between calculated
    $U$ values and should be interpreted as approximate crossovers.}
    \label{fig:Etot-vs-U-combined}
\end{figure}

 Fig.~\ref{fig:Etot-vs-U-combined} shows how the relative energies of the CPs on Ag(111) [panel (a)] and Ag(100) [panel (b)] vary with the choice of Hubbard parameter $U$ for the various phases considered. We see multiple crossings between the curves corresponding to different phases, 
and the energy-ordering of different phases is very sensitive to the choice of $U$. We also see that the energy differences resulting from a change in spin state (spin crossover) are comparable to those resulting from a change in structural 
conformation (isomerization), i.e., no single effect dominates the scale of energy differences. An external perturbation -- such as a change in temperature, pressure or exposure to light -- is likely to induce a whole slew of transitions, that involve spin crossover and/or isomerization.

For ease of reference, in Table S1 in the SI we have summarized the ordering of phases at the five values of $U$ considered. If one were to restrict one's attention to a transition on Ag(111) between the two lowest-energy phases at a given value of $U$, one sees that, depending on the value of $U$, the transition could correspond to SCO but not isomerization, isomerization but not SCO, or both SCO and isomerization. Thus, the specific phenomena associated with the lowest-energy transition are quite sensitive to the value chosen for $U$. The same is true of transitions to higher-energy phases. Further below, we will use a comparison with experimental data to choose a specific value of $U$. However, we emphasize that regardless of the value of $U$, the energy differences between different spin states are small enough that one can characterize the system as a SCO system.

We first compare the results shown in Fig.~\ref{fig:Etot-vs-U-combined} with experimentally available information to place empirical bounds on the value of $U$. Analysis of STM data has shown\cite{xu2024} that the {\it{cis}} conformation of the CP is favored on Ag(111), whereas the {\it{trans}} conformation is favored on Ag(100).
For our DFT+$U$ results (see Figs.~2 and 3), on Ag(111), the \textit{cis} conformer is lower in energy than the
\textit{trans} conformer for
$0<U<3.75$~eV. Within this interval, the lowest-energy \textit{trans}
and \textit{cis} solutions have different spin configurations for
$0.66<U<3.00$~eV. The latter boundaries are resolved more clearly in
Fig.~\ref{fig:comparing-spin-state-energies}. For example, at
$U=1$~eV, the preferred \textit{trans} solution is LS--LS--HS, whereas
the preferred \textit{cis} solution is LS--LS--LS. Thus, the experimentally observed
\textit{trans}-to-\textit{cis} structural transformation is accompanied
by a change from a mixed-spin configuration to an all-LS configuration.
On Ag(100), the experimentally observed preference for the
\textit{trans} conformer is reproduced in the DFT+$U$ results for $U>0.88$~eV. Combining the results for Ag(111) with those for Ag(100), we get $0.88<U<3.75$~eV. 

\begin{figure}[htbp]
    \centering
    \includegraphics[width=0.9\linewidth]{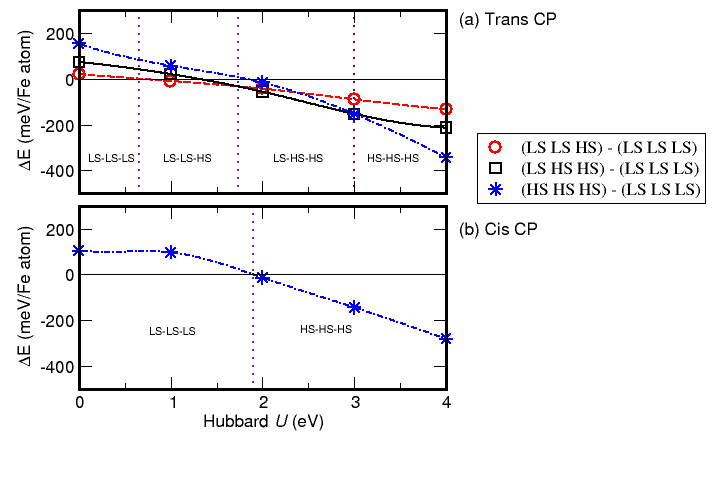}
    \caption{Relative energies of the converged spin configurations of
    the \textit{trans} and \textit{cis} chains on Ag(111). At each $U$,
    energies are referenced to the LS--LS--LS solution of the same
    conformer and normalized per Fe atom. LS--LS--HS therefore denotes
    two nominal LS Fe centers and one nominal HS Fe center, with all
    local moments aligned parallel. Dotted vertical lines mark
    interpolated changes in the lowest-energy configuration. At
    $U=1$~eV the \textit{trans} chain favours LS--LS--HS, whereas the
    \textit{cis} chain favours LS--LS--LS.}
    \label{fig:comparing-spin-state-energies}
\end{figure}

 This range of $U$ is still rather large, and within this range one sees several crossings between the curves corresponding to different phases.
To further narrow down the choice of $U$, we compare simulated and experimental STM images. In Fig.~\ref{fig:STM-comparison-U1-U3}
we compare an experimental STM image of the {\it{cis}} CP on Ag(111), with (as examples) simulated images of some of  the lowest energy {\it{cis}} CPs obtained using $U = 1$ eV and $U = 3$ eV. We note that for the former, the spin state corresponds to LS-LS-LS, whereas for the latter it is HS-HS-HS. From an examination of such images, we conclude that the best agreement between experimental and simulated STM images is obtained for $U = 1$ eV. Henceforth (unless explicitly mentioned otherwise) all DFT results presented or discussed correspond to $U = 1$ eV.

\begin{figure}[htbp]
    \centering
    \includegraphics[width=0.8\linewidth]{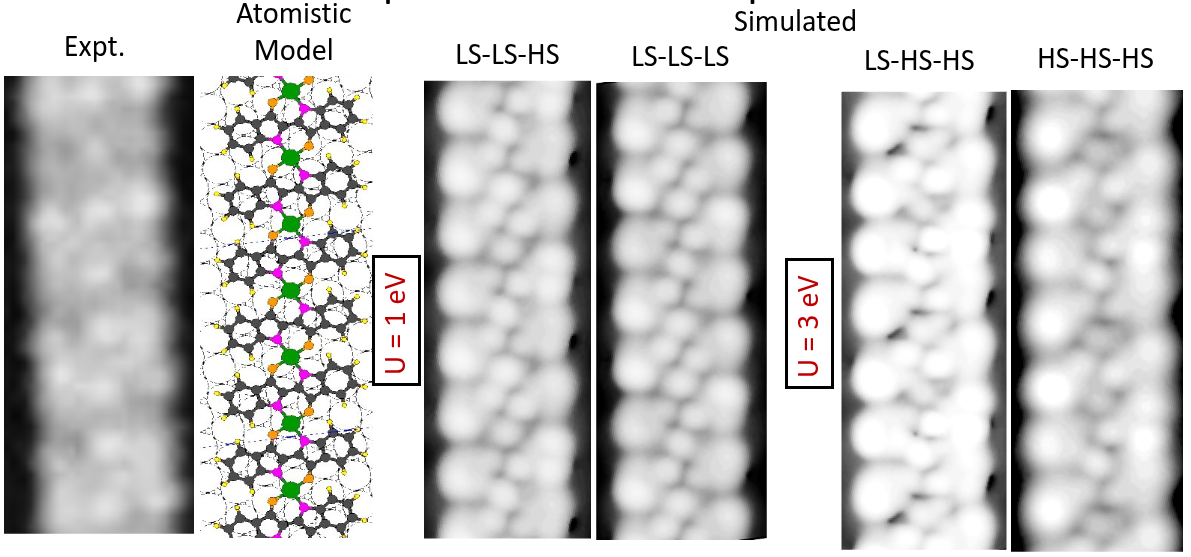}
    \caption{Comparison of experimental and simulated STM images of \textit{trans}
    CPs on Ag(111) at $U=1$~eV and $U=3$~eV, for the two lowest-energy spin
    states at each $U$. Both experimental and simulated images are obtained with the bias voltage of $-0.1$ V. The $U=1$~eV simulation matches the experimental image
    better. Black, yellow, pink, orange, green, and white spheres denote C, H,
    N, O, Fe, and Ag atoms.}
    \label{fig:STM-comparison-U1-U3}
\end{figure}

 We next discuss whether it is possible to distinguish between Fe atoms in different spin states in even non-spin-polarized STM experiments. An examination of the simulated STM images shows a slight but perceptible difference in the imaging of LS and HS Fe atoms, with the former being imaged brighter than the latter. An examination of experimental STM images (see first panel in Fig.~\ref{fig:STM-comparison-U1-U3}, as well as additional images shown in Fig.~S1 in the SI) do indeed feature some Fe atoms being imaged brighter than others, though not in any periodic fashion. This suggests that the lowest energy phase of the CPs may indeed correspond to a mixed-spin state. However, we note that we have not performed an experimental investigation of spin states, and do not have any experimental proof that Fe atoms that are imaged brighter and darker do indeed correspond to LS and HS states.

\subsection{Geometry-dependent Fe $3d$-orbital ordering in the CPs}

Fig.~5 shows the spin-dependent energy levels for the {\it{trans}} CP on Ag(111) for both the all-LS and all-HS states. Fig.~6 shows the corresponding data for the {\it{cis}} CP on Ag(111).

The supported chains no longer retain the idealized $D_{2h}$ or
$C_{2v}$ symmetry of the model systems considered above, and their Fe $3d$ states are broadened by Fe--ligand
and Fe--substrate hybridization. Consequently, Figs.~\ref{fig:Comparison-of-LS-HS-in-trans-CP}
and \ref{fig:Comparison-of-LS-HS-in-cis-CP} should not be read as bare
crystal-field eigenvalue diagrams. Instead, each horizontal line marks
the energy of the highest-intensity peak in the corresponding
spin-resolved, orbital-projected Fe $3d$ density of states, obtained using $U=1$~eV.
The SI describes this extraction explicitly and
shows the individual diagrams from which the condensed plots in Figs.~5 and 6 
were constructed.

\begin{figure}[htbp]
    \centering
    \includegraphics[width=0.8\linewidth]{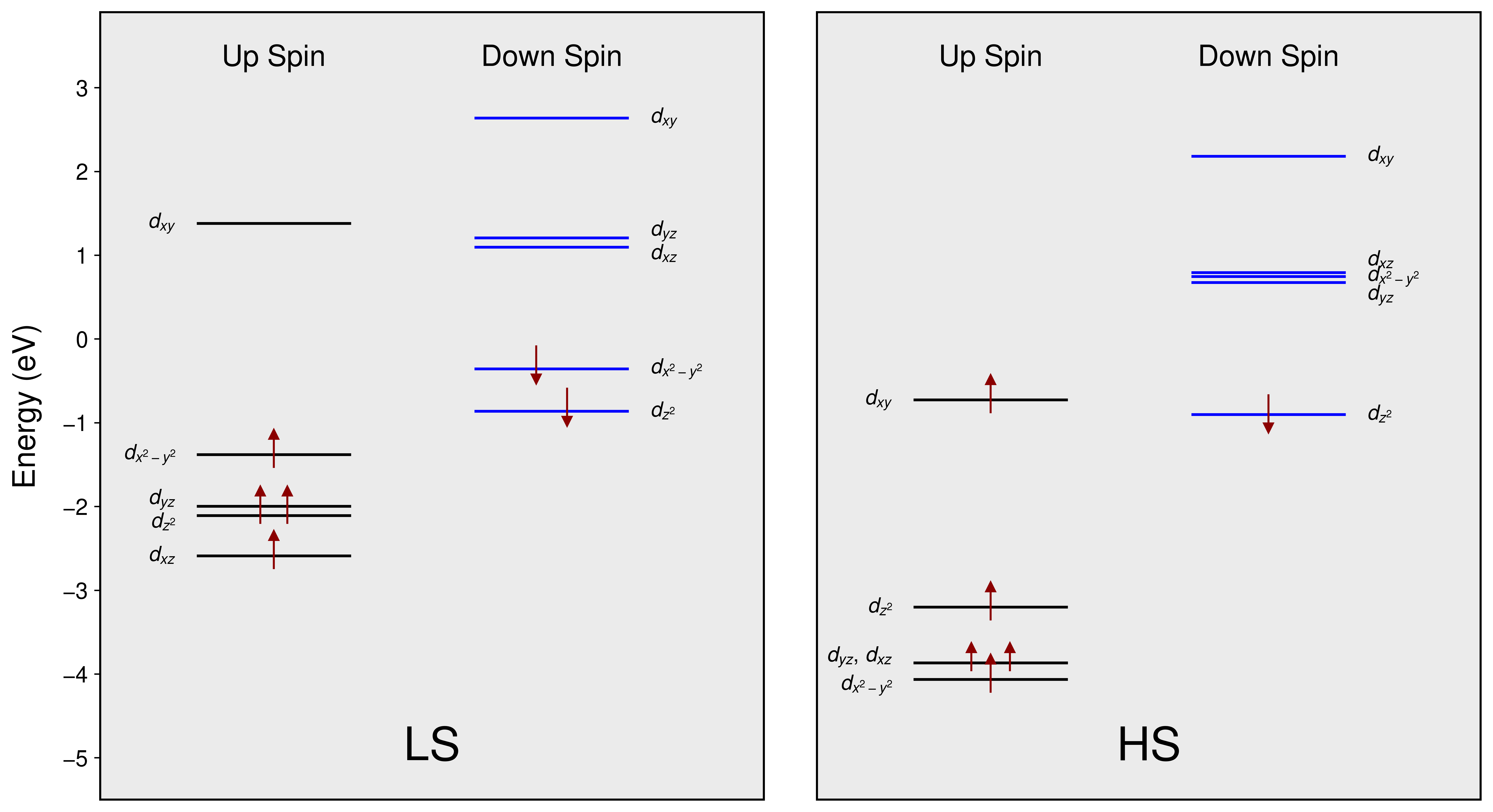}
    \caption{Spin-resolved Fe $3d$ PDOS peak positions for the
    nominal LS ($S=1$) and HS ($S=2$) solutions of the \textit{trans}
    chain on Ag(111) at $U=1$~eV. Each horizontal line is placed at the
    energy of the highest PDOS peak for the indicated orbital and spin
    channel; red arrows indicate the nominal occupation pattern of the
    converged solution. Because the states are hybridized and can be
    broadened, these peak positions are descriptive orbital-energy
    markers rather than isolated crystal-field eigenvalues.}
    \label{fig:Comparison-of-LS-HS-in-trans-CP}
\end{figure}

\begin{figure}[htbp]
    \centering
    \includegraphics[width=0.8\linewidth]{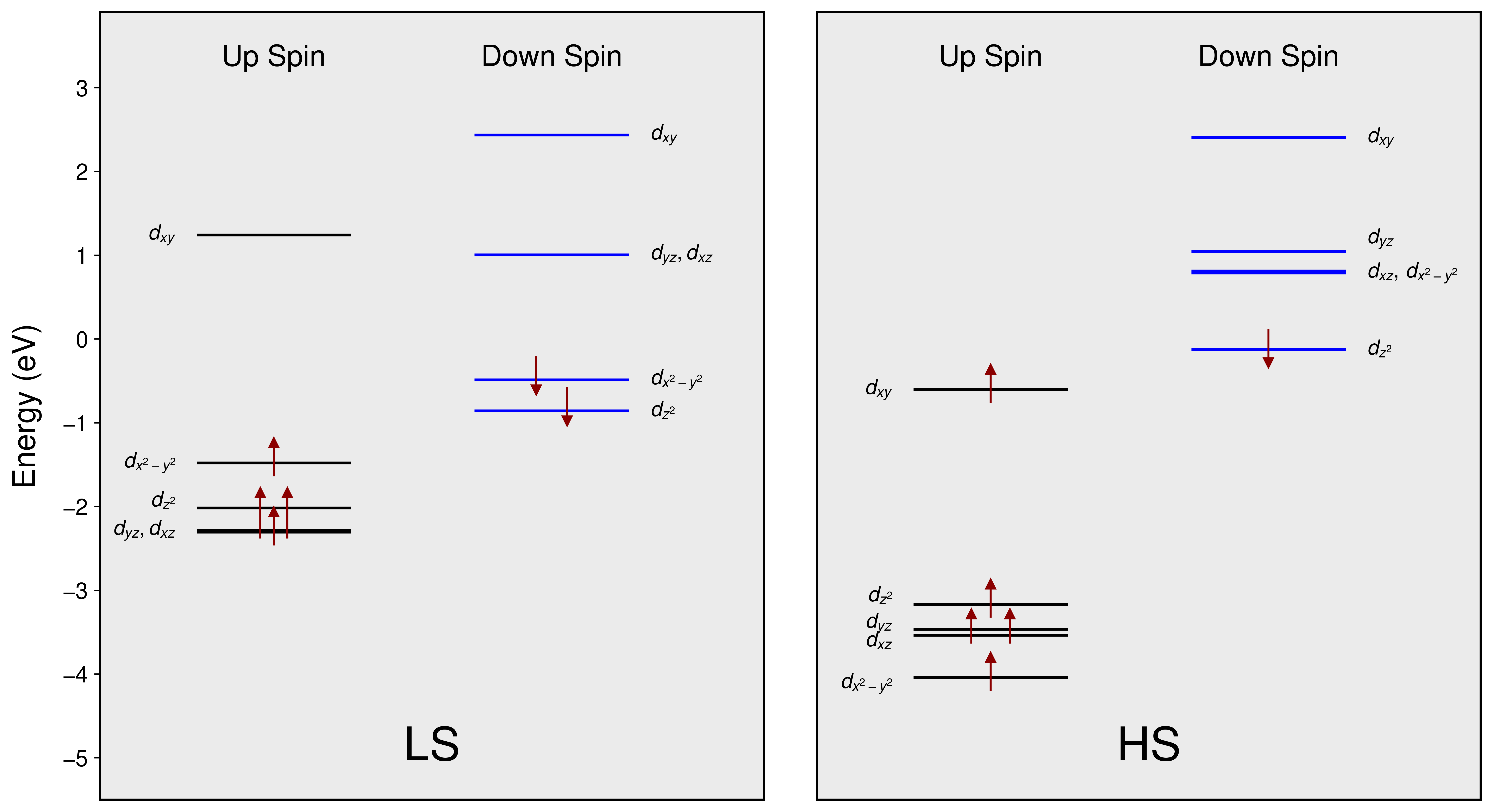}
    \caption{Spin-resolved Fe $3d$ PDOS peak positions for the
    nominal LS ($S=1$) and HS ($S=2$) solutions of the \textit{cis}
    chain on Ag(111) at $U=1$~eV. The construction is identical to
    Fig.~\ref{fig:Comparison-of-LS-HS-in-trans-CP}: horizontal lines mark
    the highest PDOS peaks and red arrows denote the nominal occupation
    pattern. The positions include ligand-field, exchange, and
    hybridization effects and are therefore not bare crystal-field
    eigenvalues.}
    \label{fig:Comparison-of-LS-HS-in-cis-CP}
\end{figure}

Comparing the two figures shows two complementary effects. First, the
change from the nominal LS to HS solution produces the expected strong
exchange rearrangement: the HS solution has five occupied
majority-spin $d$ states and only one minority-spin electron, whereas
the LS solution retains the nominal four-majority/two-minority
occupation. Second, at a fixed spin manifold the relative positions and
separations of the dominant orbital peaks are not identical in the
\textit{trans} and \textit{cis} chains. In other words, adsorption and
hybridization do not erase the geometry dependence already visible in
the idealized models. The PDOS analysis therefore supports a picture in
which isomerization modifies the detailed Fe $3d$ level hierarchy and
hence the energetic balance between the LS and HS solutions. Because a
PDOS maximum is an operational descriptor of a broadened state, we use
this comparison to interpret the trend rather than to assign unique
crystal-field eigenvalues to the supported system.}

\FloatBarrier
\subsection{Freestanding chains and strain dependence}

The surface-supported chains combine three ingredients: the intrinsic
polymer backbone, interaction with Ag, and the one-dimensional lattice constant
constraint imposed by epitaxy. Freestanding periodic chains provide a
qualitative reference in which Ag hybridization is absent while the
polymer geometry and axial lattice parameter can still be varied. They
are therefore used to determine whether the LS--HS competition is an
intrinsic property of the chain and how strongly it responds to axial
strain; they are not intended as complete models of the adsorbed
system.

\begin{figure}[htbp]
    \centering
    \includegraphics[width=0.6\linewidth]{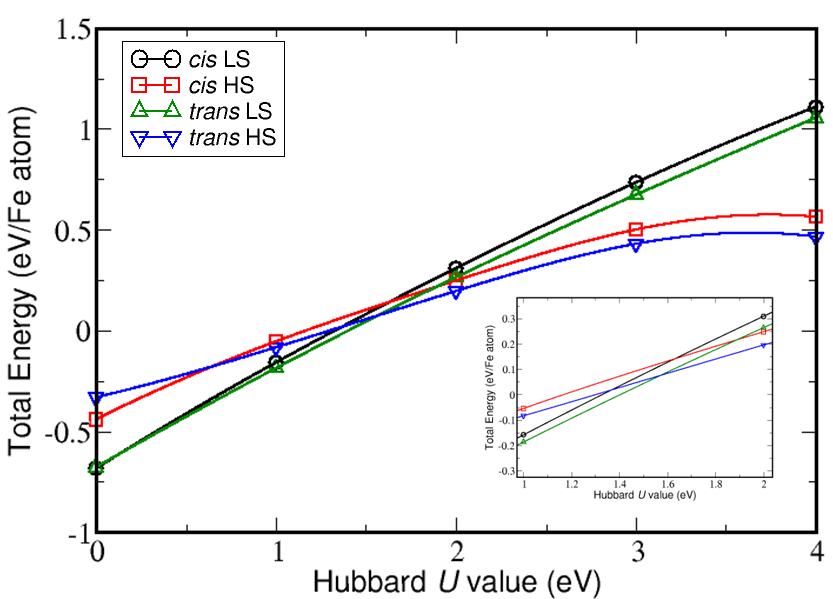}
    \caption{Total energies of isolated \textit{cis} and \textit{trans}
    chains in the ferromagnetically aligned sector as a function of $U$. The plot is used only to compare LS and HS ordering at the same
    $U$; absolute total energies at different $U$ values are not to be interpreted as a physical thermodynamic quantity. The same trend
    as in the supported systems is obtained: increasing $U$ makes the
    HS state energetically favoured over LS state. The inset shows how energetics switch for \textit{trans-cis} and LS-HS between $U$-value of 1 eV to 2 eV.}
    \label{fig:gasp_SCO_chains_FM_alignment}
\end{figure}

Figure~\ref{fig:gasp_SCO_chains_FM_alignment} shows that removal of the
Ag substrate does not remove the strong $U$ sensitivity: the lower-spin
solutions are favoured at smaller $U$, whereas the higher-spin
solutions become progressively more competitive as $U$ is increased.
As for the surface-supported chains, this behaviour is treated as a
property of the DFT+$U$ description rather than as a physical crossover
driven by $U$.

\begin{figure}[htbp]
    \centering
    \includegraphics[width=0.6\linewidth]{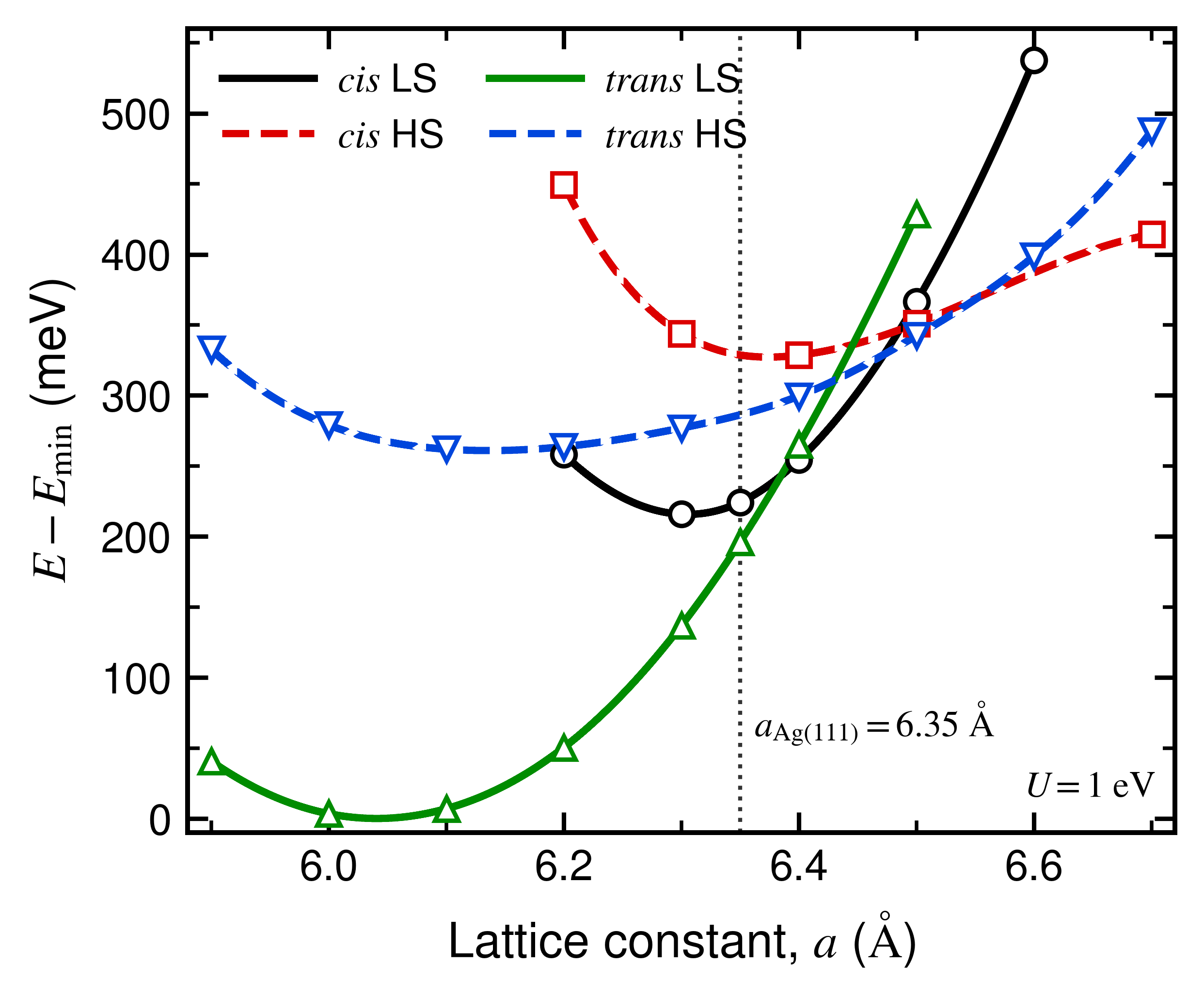}
    \caption{Relative energy, $E-E_{\min}$, as a function of chain
    lattice constant for freestanding \textit{cis} and \textit{trans}
    Fe--indigo units at $U=1$~eV in the ferromagnetically aligned sector,
    with one Fe center and one indigo ligand per unit cell. Here
    $E_{\min}$ is the minimum energy among the four plotted branches over
    the displayed lattice-constant range. LS and HS denote the nominal
    $S=1$ and $S=2$ branches. The vertical magenta line marks the
    Ag(111)-matched lattice constant, $a=6.35$~\AA.}
    \label{fig:gasp_SCO}
\end{figure}

A real structural perturbation is obtained by varying the chain lattice
constant at fixed $U=1$~eV. Figure~\ref{fig:gasp_SCO} shows the resulting
relative energies. The \textit{trans} LS branch is the lowest of the
four branches over the explored range and has its minimum near
$a\simeq6.0$--6.1~\AA. Its HS branch is substantially flatter and becomes
relatively more favourable on expansion. The \textit{cis} LS and HS
branches lie closer together near the Ag(111)-matched lattice constant,
showing that the spin-state balance in this geometry is particularly
sensitive to the imposed axial constraint.

Unlike the formal variation of $U$, this lattice-constant dependence is
a genuine magnetoelastic response: expansion lowers the HS branch
relative to the LS branch. The result establishes strain as an
additional physical contribution to the LS--HS competition. It is also
noteworthy that the global minimum of the freestanding calculations is
the \textit{trans} LS branch, whereas the supported Ag(111) calculations
favour the \textit{cis} conformer over the relevant $U$ interval. This
contrast shows that the Ag-supported system cannot be understood from
the isolated polymer backbone alone; substrate interaction and the
substrate-imposed lattice constraint materially modify the energetic
balance.

\section{Conclusion}

We have examined the spin-state energetics of one-dimensional Fe--indigo
coordination polymers on Ag(111) and Ag(100) using DFT+$U$. On Ag(111), the
preferred \textit{trans} and \textit{cis} solutions differ in spin
configuration over the interpolated interval $0.66 < U < 3.00$~eV; at the
reference value $U = 1$~eV, the \textit{trans} chain favors the mixed
LS--LS--HS configuration, while the \textit{cis} chain favors LS--LS--LS.
The independently calculated conformer ordering reproduces the
experimentally observed preference for \textit{cis} chains on Ag(111) and
\textit{trans} chains on Ag(100) over $0.88 < U < 3.75$~eV. The overlap of
these two intervals, $0.88 < U < 3.00$~eV, therefore simultaneously
reproduces the surface-dependent conformer ordering and the
isomer-dependent spin configuration, and contains the reference value
$U = 1$~eV used throughout. The precise interpolated boundaries should not
be read as physical transition values of $U$; rather, the scan establishes
that this fixed-$U$ contrast between the two isomers persists over a
finite parameter range.

A recurring theme of this work is that structural isomerization and spin
crossover are not independent phenomena in this system, but are coupled
through the ligand field at the Fe centers. Each combination of
conformation and spin configuration can be regarded as a distinct phase of
the coordination polymer, and we find that the energy differences
separating these phases are of comparable magnitude regardless of whether
the underlying change is a spin crossover, an isomerization, or both
together. Consequently, an external stimulus such as a change in
temperature, pressure, or illumination is likely to drive transitions that
combine both effects rather than either in isolation.

Idealized $D_{2h}$ and $C_{2v}$ coordination models show a pronounced
reordering of the Fe $3d$ manifold when the donor geometry changes,
including an upward displacement of $d_{xz}$ from the bottom of the
\textit{trans} manifold to the vicinity of $d_{yz}$ in the \textit{cis}
model. In the supported chains, the dominant spin-resolved PDOS peak
positions retain this geometry dependence, although hybridization with the
substrate means that these peaks should be read as descriptors of a
broadened state rather than as bare crystal-field eigenvalues.
Freestanding-chain calculations add a further, distinct contribution:
lattice expansion stabilizes the HS branch relative to the LS branch,
establishing strain as a genuine physical driver of the spin-state
balance, in contrast to the purely methodological sensitivity associated
with the Hubbard $U$ scan.

Taken together, these calculations support a microscopic picture in which
isomerization reorganizes the Fe coordination field, while substrate
interaction and lattice constraint further renormalize the resulting
spin-state balance. This picture is deliberately limited to the
ferromagnetically aligned, collinear sector sampled here; antiferromagnetic
or other non-collinear intersite arrangements were not surveyed
exhaustively and could alter the absolute magnetic ground state. Within
the scope of the present calculations, however, the isomer-dependent
change in the preferred local spin configuration is robust over the
parameter window identified above, and establishes a microscopic basis for
the interplay between isomerization and spin crossover in this
one-dimensional coordination polymer.

\begin{suppinfo}
Supporting Information contains the procedure used to extract the
dominant spin-resolved Fe $3d$ PDOS peak positions used in
Figs.~\ref{fig:Comparison-of-LS-HS-in-trans-CP} and
\ref{fig:Comparison-of-LS-HS-in-cis-CP}, together with the individual
LS and HS level diagrams for the \textit{trans} and \textit{cis}
coordination polymers on Ag(111) at $U=1$~eV (PDF).
\end{suppinfo}

\begin{acknowledgement}
  Support and resources were provided by the PARAM Yukti, National Supercomputing Mission and Central Computing Facility, both at JNCASR, Bangalore, India. R.C. thanks the UGC, Government of India, for a doctoral fellowship. H.X. thanks the China Scholarship Council (CSC) for a doctoral scholarship. S.N. also acknowledges the Sheikh Saqr Laboratory of ICMS, JNCASR and the Anna Boyksen Fellowship of the Technical University of Munich, Institute for Advanced Study. 
\end{acknowledgement}

\section*{Data Availability Statement}
All data are available from the authors upon reasonable request.

\section*{Conflicts of Interest}
The authors declare no competing financial interest.

\bibliography{refer}

\newpage
\section*{Supplementary Information}

\begin{table}[t]
\centering
\caption{Relative energetic ordering of the different isomeric and spin-state
configurations as a function of the Hubbard parameter $U$.
The configurations are arranged from higher to lower energy.}
\label{tab:state_ordering_U}
\renewcommand{\arraystretch}{1.25}
\setlength{\tabcolsep}{4pt}

\resizebox{\textwidth}{!}{%
\footnotesize
\begin{tabular}{c|ccccc}
\hline
 & $U=0$ & $U=1$ & $U=2$ & $U=3$ & $U=4$ \\
\hline

1 &
\textit{trans} HS--HS--HS &
\textit{trans} HS--HS--HS &
\textit{trans} LS--LS--LS &
\textit{trans} LS--LS--LS &
\textit{trans} LS--LS--LS \\

2 &
\textit{cis} HS--HS--HS &
\textit{trans} LS--HS--HS &
\textit{trans} HS--HS--HS &
\textit{cis} LS--LS--LS &
\textit{cis} LS--LS--LS \\

3 &
\textit{trans} LS--HS--HS &
\textit{cis} HS--HS--HS &
\textit{trans} LS--LS--HS &
\textit{trans} LS--LS--HS &
\textit{trans} LS--LS--HS \\

4 &
\textit{trans} LS--LS--HS &
\textit{trans} LS--LS--LS &
\textit{trans} LS--HS--HS &
\textit{trans} LS--HS--HS &
\textit{trans} LS--HS--HS \\

5 &
\textit{trans} LS--LS--LS &
\textit{trans} LS--LS--HS &
\textit{cis} LS--LS--LS &
\textit{trans} HS--HS--HS &
\textit{cis} HS--HS--HS \\

6 &
\textbf{\textit{cis} LS--LS--LS} &
\textbf{\textit{cis} LS--LS--LS} &
\textbf{\textit{cis} HS--HS--HS} &
\textbf{\textit{cis} HS--HS--HS} &
\textbf{\textit{trans} HS--HS--HS} \\

\hline
\multicolumn{1}{c|}{\textbf{Lowest-energy state}} &
\textbf{\textit{cis} LS--LS--LS} &
\textbf{\textit{cis} LS--LS--LS} &
\textbf{\textit{cis} HS--HS--HS} &
\textbf{\textit{cis} HS--HS--HS} &
\textbf{\textit{trans} HS--HS--HS} \\
\hline

\end{tabular}%
}
\end{table}


\begin{figure}[htbp]
    \centering
    \begin{subfigure}[b]{0.53\linewidth}
        \centering
        \includegraphics[width=\linewidth]{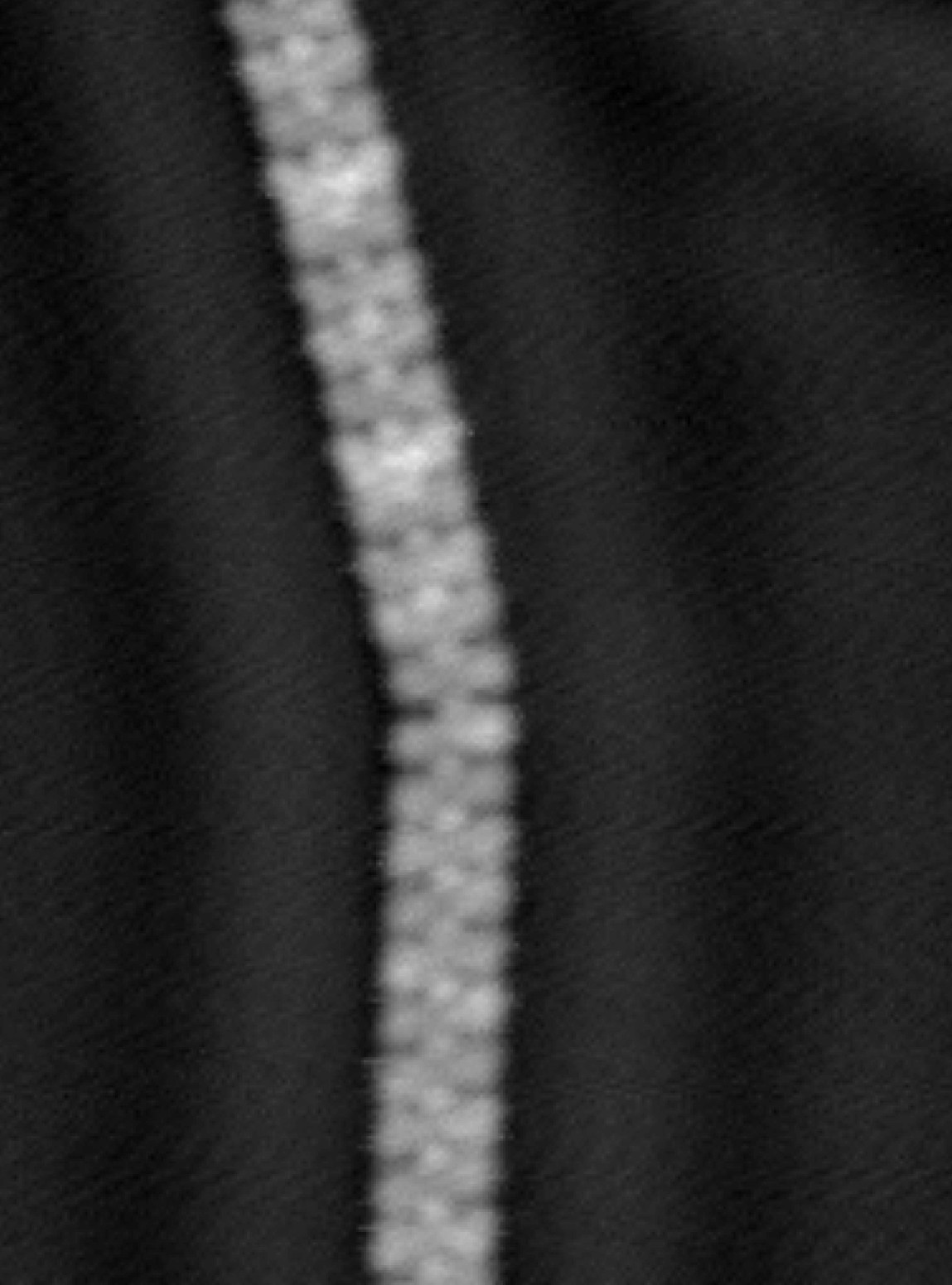}
        \caption{}
        \label{fig:Expt-STM-image-1}
    \end{subfigure}
    \hfill
    \begin{subfigure}[b]{0.38\linewidth}
        \centering
        \includegraphics[width=\linewidth]{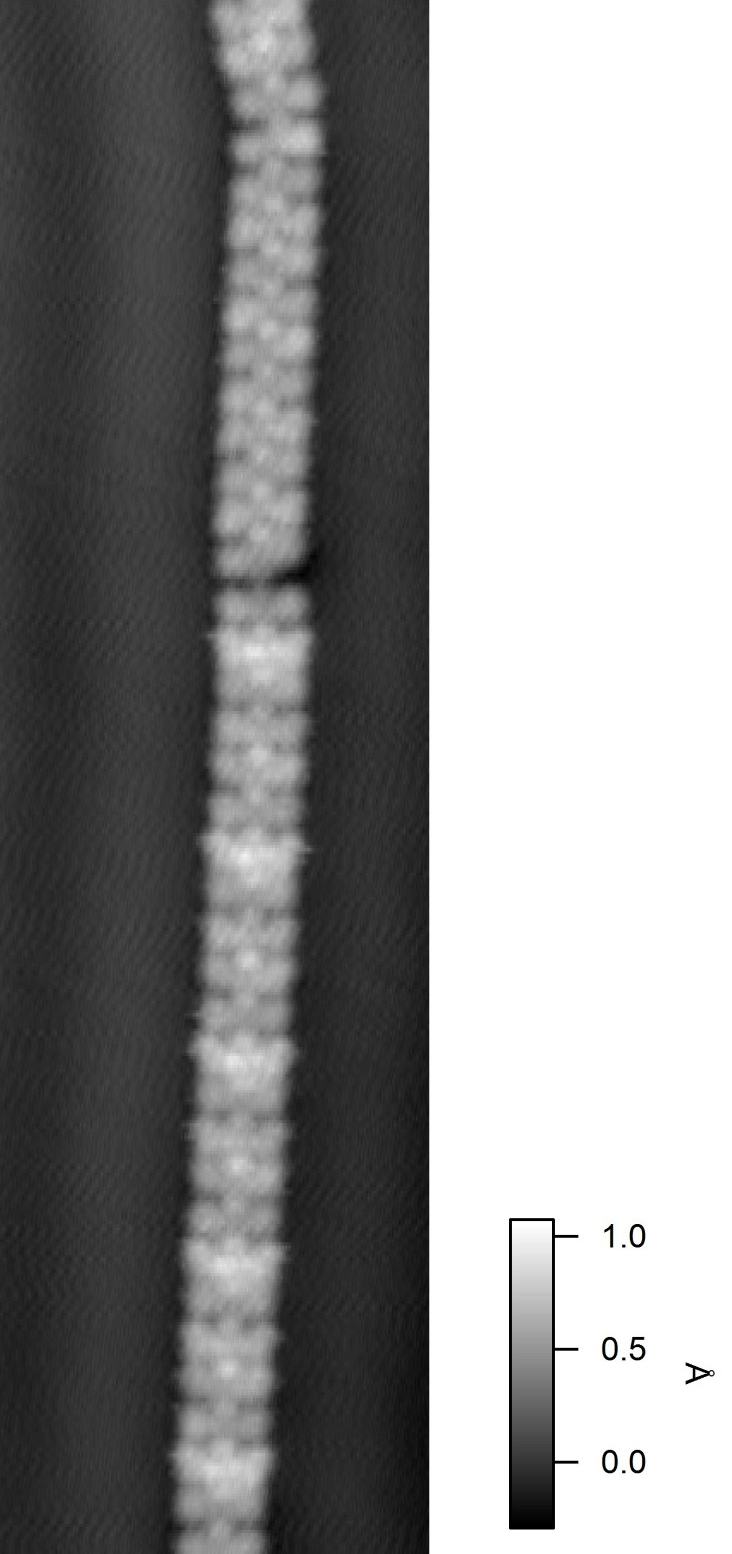}
        \caption{}
        \label{fig:Expt-STM-image-2}
    \end{subfigure}
    \caption{(a) Representative constant-current STM image of an extended indigo–Fe CP acquired at \(V=+0.10\) V. The chain contains both \textit{cis} and \textit{trans} CP segments and exhibits nonuniform apparent-height contrast at the Fe sites. (b,c) STM images showing different regions of the same chain, acquired under the same imaging conditions. Several Fe centers appear slightly brighter or darker than neighboring sites, without any obvious periodic modulation of the contrast. The simulated STM images likewise predict a small spin-state-dependent contrast, with LS Fe centers appearing brighter than HS Fe centers (see Fig.~4 of the main manuscript). The experimentally observed contrast variations are therefore consistent with the possible coexistence of different Fe spin states within the chain. However, because the measurements are non-spin-polarized, individual bright and dark Fe sites cannot be unambiguously assigned to LS and HS states, respectively. Grayscale bars indicate the apparent-height scale.}
    \label{fig:Expt-STM-images}
\end{figure}

\section{Toy coordination models for the \textit{trans} and \textit{cis} Fe environments}

\begin{figure}[htbp]
    \centering
    \includegraphics[width=0.8\linewidth]{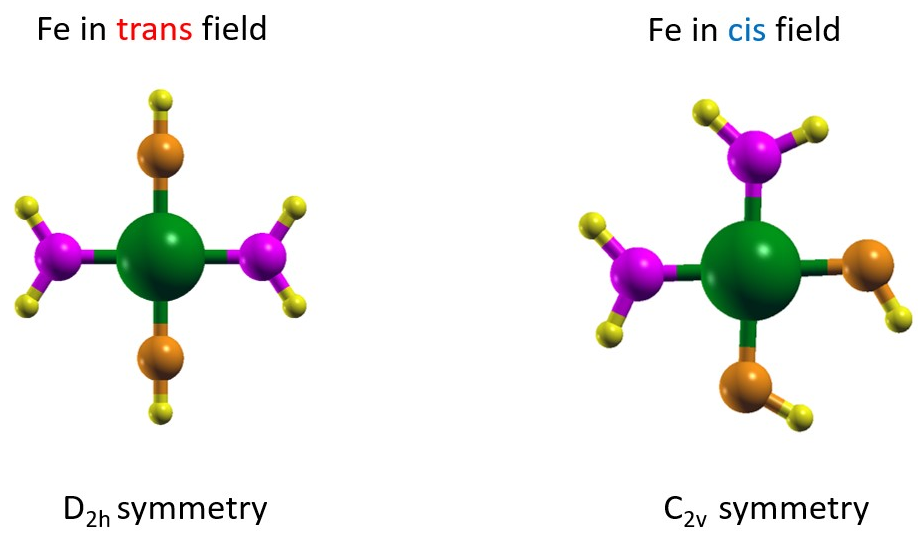}
    \caption{Idealized coordination models used to compare the Fe
    $d$-orbital ordering in the \textit{trans} and \textit{cis}
    environments. The $x$ and $y$ axes lie in the Fe coordination plane
    and $z$ is normal to that plane. Yellow, pink, orange, and green
    spheres denote H, N, O, and Fe, respectively.}
    \label{fig:SI-Toy-models-for-cis-trans}
\end{figure}

To understand the Fe $d$-orbital ordering, we first built simple, freestanding
models: an Fe centre bonded to two N and two O atoms, with the remaining
valences capped by H (Fig.~\ref{fig:SI-Toy-models-for-cis-trans}). The $x$ and
$y$ axes lie in the Fe coordination plane and $z$ is perpendicular to it; we
use this convention for all orbital labels below.

\begin{figure}[htbp]
    \centering
    \begin{subfigure}[b]{0.50\linewidth}
        \centering
        \includegraphics[width=\linewidth]{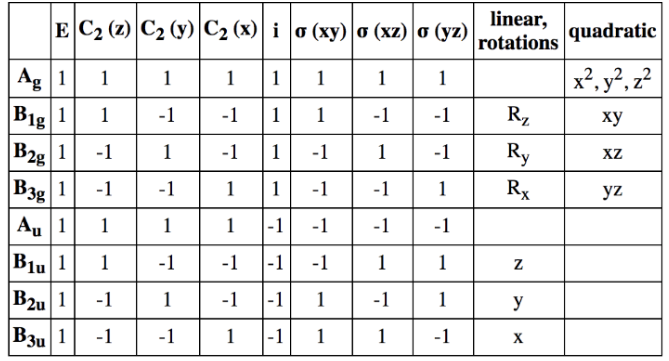}
        \caption{$D_{2h}$}
        \label{fig:D2h-character-table}
    \end{subfigure}
    \hfill
    \begin{subfigure}[b]{0.44\linewidth}
        \centering
        \includegraphics[width=\linewidth]{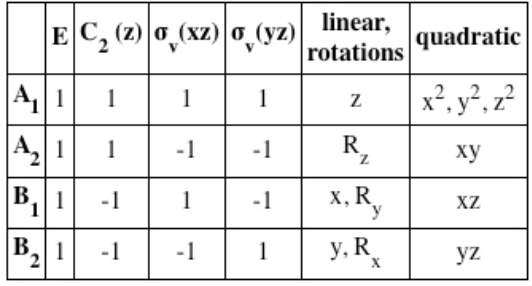}
        \caption{$C_{2v}$}
        \label{fig:c2v-character-table}
    \end{subfigure}
    \caption{Character table of $D_{2h}$ and $C_{2v}$ point group. This character table clearly shows that none of the quadratic functions corresponding to $d$-orbitals will be degenerate. }
    \label{fig:Etot-vs-U-combined}
\end{figure}

\begin{figure}[htbp]
    \centering
    \includegraphics[width=0.7\linewidth]{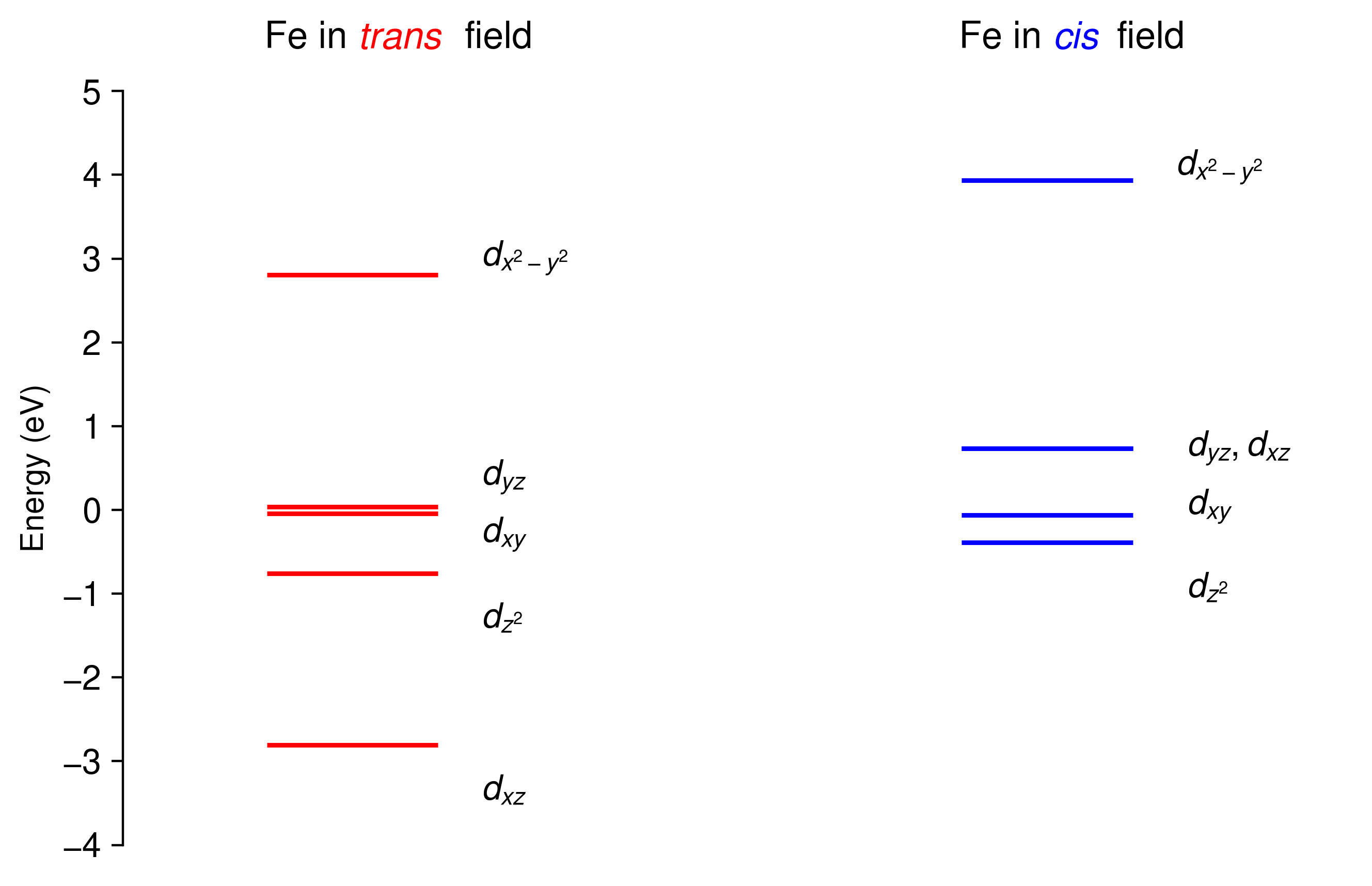}
    \caption{Fe $d$-orbital energies from non-spin-polarized calculations
    of the idealized models in Fig.~\ref{fig:SI-Toy-models-for-cis-trans},
    plotted relative to the Fermi level ($E_F=0$). There are no
    symmetry-enforced degeneracies, although nearly coincident levels
    occur. The ordering of the lower $d$ manifold differs markedly
    between the two coordination geometries, while $d_{x^2-y^2}$
    remains the highest level in both.}
    \label{fig:SI-Fe-splitting-system-in-trans-cis-field}
\end{figure}

The idealized \textit{trans} and \textit{cis} environments have $D_{2h}$ and
$C_{2v}$ symmetry. Neither symmetry forces any $d$-orbitals to be degenerate,
though some can end up close in energy. Figure~\ref{fig:SI-Fe-splitting-system-in-trans-cis-field}
shows the resulting orbital energies. In both geometries, $d_{x^2-y^2}$ sits
highest, as expected for a four-coordinate, near-square-planar Fe centre. The
lower orbitals, however, reorder on isomerisation: $d_{xz}$ is lowest in the
\textit{trans} field, but shifts up to sit near $d_{yz}$ in the \textit{cis}
field. So changing the donor arrangement reorders the orbitals -- it does not
just shift them all together.

This high-lying $d_{x^2-y^2}$ level is also why we call the lower-spin state
$S=1$, not $S=0$ as in typical octahedral Fe(II) complexes: populating
$d_{x^2-y^2}$ costs too much energy, but the remaining four orbitals are close
enough in energy that one of them stays singly occupied rather than all six
electrons pairing up in three. Square-planar Fe(II) complexes are known to
behave this way (Ref.~34 of the main text: Pascualini \textit{et al.},
\textit{Chem. Sci.} \textbf{2015}, \textit{6}, 608--612).

\section{Extraction of Fe \texorpdfstring{$3d$}{3d} orbital-energy markers from the projected density of states}

The level diagrams in the main text are built directly from the spin- and
orbital-resolved PDOS of the Fe $3d$ states. They give a compact picture of
the dominant orbital energies in the hybridized, surface-supported chains --
not exact crystal-field eigenvalues.

\begin{figure}[htbp]
    \centering
    \includegraphics[width=0.95\linewidth]{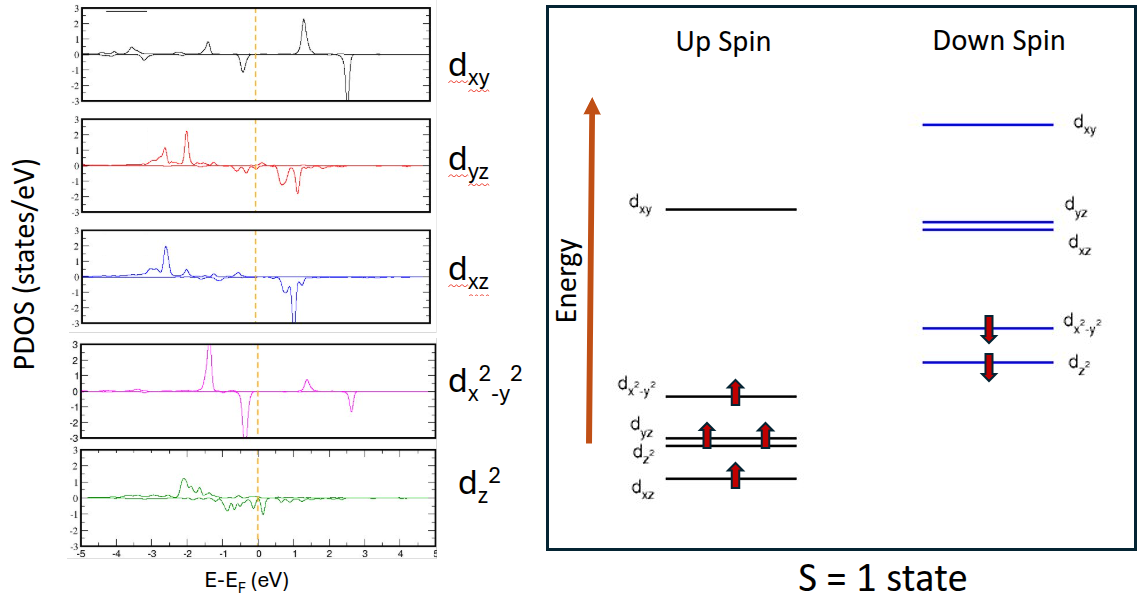}
    \caption{Nominal LS ($S=1$) Fe centre in the \textit{trans} coordination
    polymer on Ag(111) at $U=1$~eV. Horizontal lines mark the
    highest-intensity peaks of the spin- and orbital-resolved Fe $3d$ PDOS;
    red arrows show the nominal LS occupation.}
    \label{fig:SI-trans-LS}
\end{figure}

\begin{figure}[htbp]
    \centering
    \includegraphics[width=0.95\linewidth]{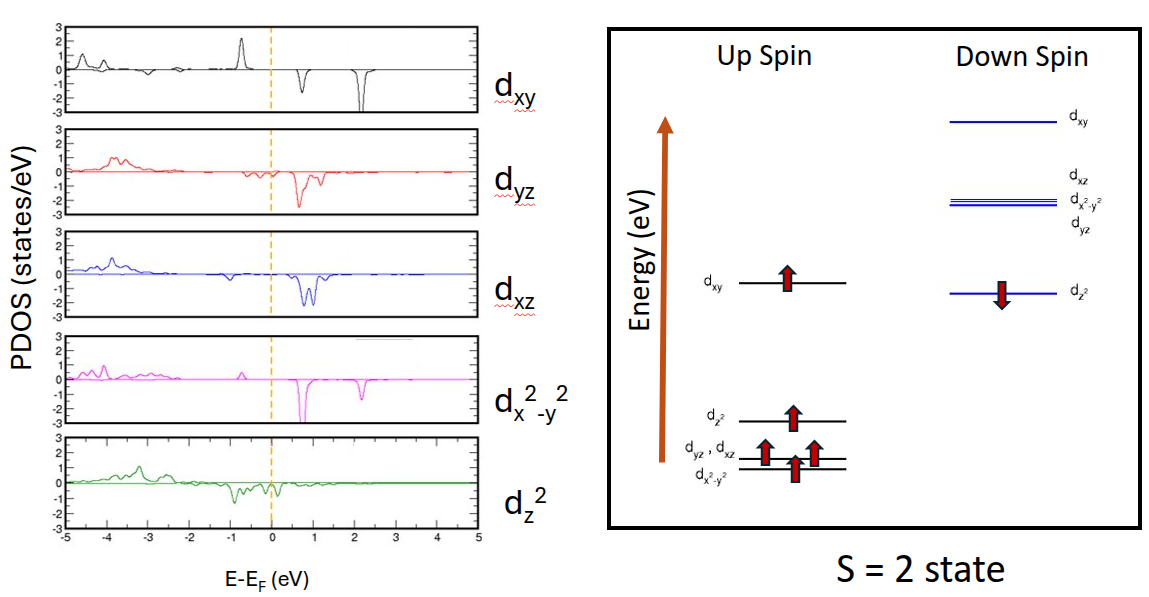}
    \caption{Nominal HS ($S=2$) Fe centre in the \textit{trans} coordination
    polymer on Ag(111) at $U=1$~eV. Same construction as Fig.~\ref{fig:SI-trans-LS}.}
    \label{fig:SI-trans-HS}
\end{figure}

\begin{figure}[htbp]
    \centering
    \includegraphics[width=0.95\linewidth]{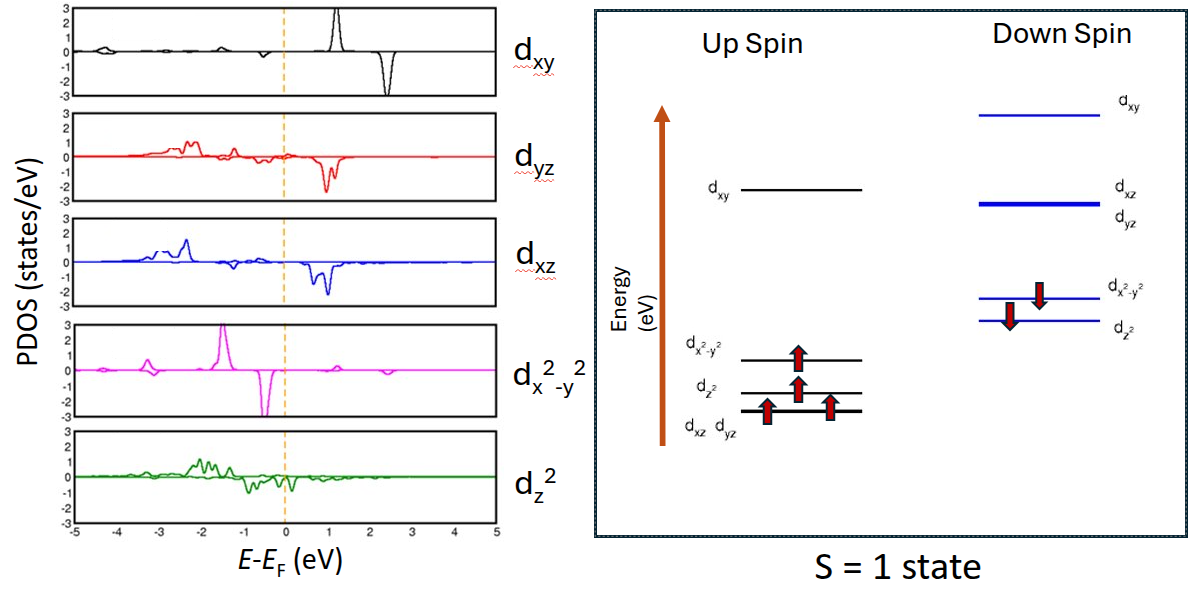}
    \caption{Nominal LS ($S=1$) Fe centre in the \textit{cis} coordination
    polymer on Ag(111) at $U=1$~eV. Same construction as Fig.~\ref{fig:SI-trans-LS}.}
    \label{fig:SI-cis-LS}
\end{figure}

\begin{figure}[htbp]
    \centering
    \includegraphics[width=0.95\linewidth]{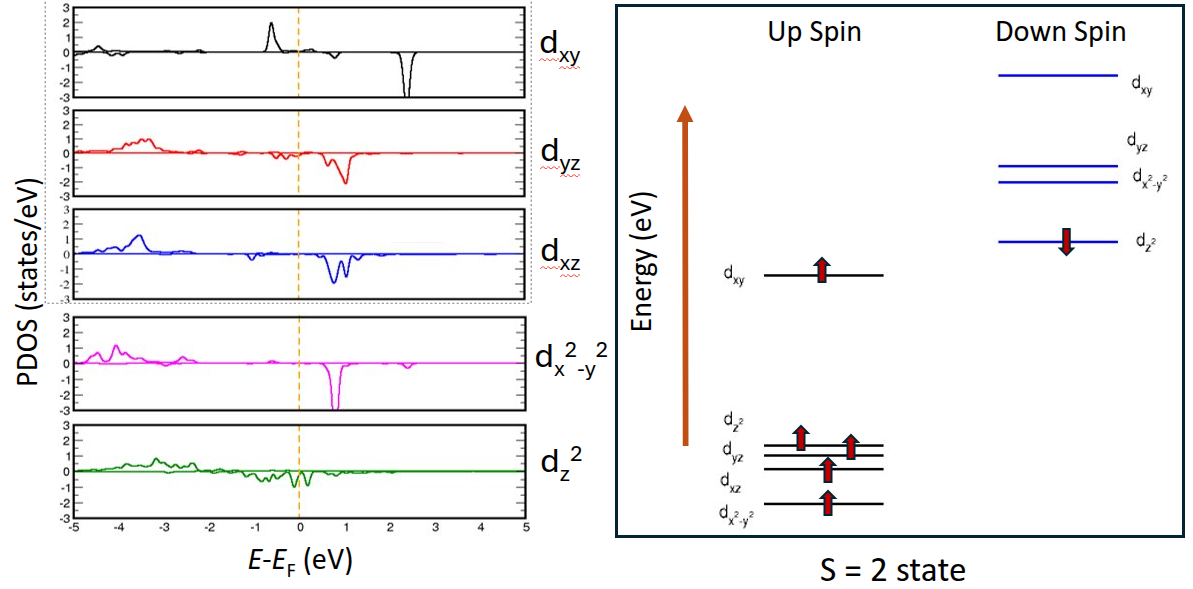}
    \caption{Nominal HS ($S=2$) Fe centre in the \textit{cis} coordination
    polymer on Ag(111) at $U=1$~eV. Same construction as Fig.~\ref{fig:SI-trans-LS}.}
    \label{fig:SI-cis-HS}
\end{figure}

For each Fe $3d$ orbital $m \in \{d_{xy}, d_{xz}, d_{yz}, d_{z^2},
d_{x^2-y^2}\}$ and each spin channel $\sigma$, we take the energy of that
orbital's tallest PDOS peak,
\begin{equation}
E^{\rm peak}_{m\sigma} = \arg\max_{E} D_{m\sigma}(E),
\end{equation}
within the Fe $3d$ energy range. The horizontal lines in
Figs.~\ref{fig:SI-trans-LS}--\ref{fig:SI-cis-HS} mark these peak positions
directly; no fitting or averaging was applied.

This suits the present system well, since adsorption on Ag(111) breaks the
idealized molecular symmetries and adds Fe--ligand and Fe--substrate
hybridization, which can broaden each orbital's PDOS or add extra structure.
So $E^{\rm peak}_{m\sigma}$ marks where most of that orbital's spectral
weight sits, not an exact crystal-field level. We use it to compare orbital
ordering and spacing between the \textit{trans} and \textit{cis}
environments, and between the LS and HS solutions.

Red arrows in Figs.~\ref{fig:SI-trans-LS}--\ref{fig:SI-cis-HS} show the
nominal occupation: four majority-spin and two minority-spin electrons for
LS ($S=1$), or five majority-spin and one minority-spin electron for HS
($S=2$). These are a guide to the spin state, not exact orbital occupations.

The four diagrams below are the individual LS/HS, \textit{trans}/\textit{cis}
panels behind the condensed comparisons in the main text, all built the same
way so the peak positions can be compared directly.

\end{document}